\documentclass[prl,twocolumn,showpacs,superscriptaddress,tightenlines]{revtex4}

\usepackage{graphicx} 
\usepackage{amsmath}
 
\begin{document}

\title{Induced Interactions for Ultracold Fermi Gases in Optical Lattices}
\author{D.-H. Kim}
\affiliation{Department of Applied Physics, Helsinki University of Technology, P.O. Box 5100, 02015 HUT, Finland}
\author{P. T\"{o}rm\"{a}}
\affiliation{Department of Applied Physics, Helsinki University of Technology, P.O. Box 5100, 02015 HUT, Finland}
\author{J.-P. Martikainen}
\affiliation{NORDITA, 106 91 Stockholm, Sweden}

\date{\today}

\begin{abstract}
We investigate the effect of optical lattices on the BCS superfluidity by using 
the Gorkov--Melik-Barkhudarov (GMB) correction 
for a two-component Fermi gas. We find that the suppression of the order 
parameter is strongly enhanced by the lattice effects. 
The predictions made by the GMB corrections are in qualitative and, 
for the cases studied, in quantitative agreement with previous
quantum Monte Carlo results. We discuss how the GMB correction extends 
the validity of the mean-field theory to a wider range of tunable optical 
lattice systems in different dimensions.
\end{abstract}


\maketitle

Bardeen, Cooper, and Schrieffer (BCS) explained superconductivity
by the condensation of fermion pairs in the presence of arbitrarily attractive 
interaction~\cite{Bardeen1957a}. Based on the idea of the BCS pairing, 
in dilute Fermi gases, the critical temperature was derived in terms of 
scattering length~\cite{SadeMelo1993a,Stoof1996b}.  
Gorkov and Melik-Barkhudarov (GMB) extended this calculation by incorporating 
many-body effects, which turned out to reduce the critical temperature 
by a factor $(4e)^{1/4} \approx 2.22$~\cite{Gorkov1961a}.  
Fermionic superfluidity has recently attracted renewed attention 
in connection with the realization of ultra-cold atomic gases that 
allow direct observation of quantum many-body phenomena 
in highly controllable 
environments~\cite{Lewenstein2007a,Bloch2008a,Giorgini2008a}. 
In particular, optical lattices are a perfect platform for emulating
crystalline structures of superconductors. 
While indirect evidence of superfluidity in a system with an optical lattices
has been recently reported~\cite{Chin2006a},  
the full characterization of fermionic superfluidity and strongly
correlated quantum states in optical lattices 
is still under active study, both theoretical and experimental
\cite{Jordens2008a,Schneider2008a}.
In this Letter we focus on how the lattice potential influences the BCS-type 
superfluid transition by employing the GMB correction. 

We calculate the mean-field BCS  order parameters at zero temperature 
in three and two dimensional 
(3D and 2D) lattices with various settings including the crossover from 3D 
to 1D. In all the ranges of lattice parameters examined, 
we find that the induced interaction introduced 
by the correction leads to remarkable reduction in the order parameter 
from the usual BCS result. This deviation turns out to be much more 
pronounced in the lattices than in homogeneous gases and becomes 
increasingly significant at higher fillings. 
In particular, in 2D, we find quantitative agreement with previous 
quantum Monte Carlo (QMC) calculations for the cases studied. 
Furthermore, near half filling in 2D, 
the rapid decreasing behavior of the order parameter is in qualitative 
agreement with the QMC predictions. At half filling in 2D, 
the induced interaction diverges because of Fermi surface nesting.  
This divergence is connected to the signature of the charge density waves, 
known to coexist with superfluidity at half filling in 2D.  

We consider a system composed of two different fermionic species 
denoted by $\uparrow$ and $\downarrow$. Each component is in a lattice
with adjustable tunneling strengths $t_{\downarrow\alpha}$ and 
$t_{\uparrow\alpha}$ in direction $\alpha\in\{x,y,z\}$. 
When the lattice potential is sufficiently deep so that we can consider only 
nearest-neighbor tunnelings and on-site interactions, the system is 
described by the Hubbard Hamiltonian   
$\mathcal{H} = 
-\sum_{\sigma,\alpha} \sum_{i_\alpha} t_{\sigma \alpha} 
\mathcal{K}_{\sigma i_\alpha} 
+U_0\sum_{\mathbf{i}} \hat{n}_{\uparrow \mathbf{i}}
\hat{n}_{\downarrow \mathbf{i}}
-\mu \sum_{\sigma,\mathbf{i}} \hat{n}_{\sigma \mathbf{i}}$ ,
where 
$\mathcal{K}_{\sigma i}\equiv
\hat{\psi}^\dagger_{\sigma i+1}\hat{\psi}_{\sigma i}
+ \hat{\psi}^\dagger_{\sigma i}\hat{\psi}_{\sigma i+1}$, 
and ${\hat \psi}_{\sigma,{\bf i}}$(${\hat \psi}_{\sigma,{\bf i}}^\dagger$) 
is the annihilation(creation) operator for atoms of type $\sigma$ 
at a site ${\bf i}\equiv(i_x,i_y,i_z)$.
The chemical potential and density operator are denoted by $\mu$
and ${\hat n}_{\sigma,{\bf i}}$, respectively. We consider negative interaction 
strengths $U_0$. 
For noninteracting gases, the above Hamiltonian is diagonalized with
dispersion
$\xi_\sigma({\bf k})=
2\sum_\alpha t_{\sigma\alpha}\left[1-\cos(k_\alpha)\right] -\mu$,
where the lattice spacing is chosen to be unity.
In the weak coupling regime, using the standard BCS theory with this
dispersion and the interaction $U_0$, without many-body
corrections, one recovers the usual BCS prediction for the critical 
temperature in the long wavelength limit. 
However, because the scattering length $a$ gives exponential contribution 
to the critical temperature as $T_c \propto \exp(-\pi/k_F|a|)$, 
where $k_F$ is the Fermi momentum, 
a small correction to the interaction term $k_Fa$ can considerably 
change $T_c$ even in the weak coupling regime. 
For instance, a second-order correction $\delta$ in 
$k_F|a| \to k_F|a|\left(1+\delta k_F|a|\right)$ leads to $T_c \to e^{\delta} T_c$.

\begin{figure}
\includegraphics[width=0.47\textwidth]{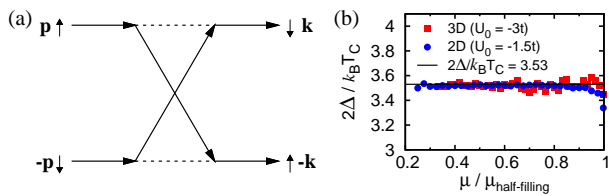} 
\caption{
(a) The diagram representing the induced interaction
$U_\mathrm{ind}({\bf p},{\bf k})$. Arrowed and dashed lines describe 
fermionic propagators and the coupling $U_0$ between the atoms.
(b) Equivalence between the zero-temperature order parameter $\Delta$ and 
the critical temperature $T_c$ in the weak coupling regime.   
}
\label{fig1}
\end{figure}

For a two component Fermi gas with an $s$-wave interaction between 
components, the relevant second order correction to the effective 
interaction is represented by the diagram in Fig.~\ref{fig1}(a), 
which describes the exchange of density and 
spin fluctuations ~\cite{Heiselberg2000a,Baranov2008a}. 
The diagram leads to the induced interaction term, which can be 
derived for an infinite-size system in $D$ dimensions as
\begin{equation}
U_\mathrm{ind}({\bf p},{\bf k})=
-U_0^2\int \frac{d\mathbf{q}}{(2\pi)^D}
\frac{f_{\uparrow,\mathbf{p}+\mathbf{k}+\mathbf{q}}-f_{\downarrow,\mathbf{q}}}
{\xi_\uparrow(\mathbf{p}+\mathbf{k}+\mathbf{q}) - \xi_\downarrow(\mathbf{q})} ,
\label{eq:induced}
\end{equation}
where the Fermi distribution 
$f_{\sigma,\mathbf{k}}=1/\left[1+\exp(\beta\xi_\sigma(\mathbf{k}))\right]$,
with $\beta=1/k_BT$. 
There are two noticeable properties in Eq. (\ref{eq:induced}). 
First, $U_\mathrm{ind}$ is always positive. Thus, this correction
screens the negative interatomic potential, which consequently reduces 
the critical temperature. Second, the static Lindhard function, representing 
the spin or density susceptibility $\chi_0$ of a noninteracting gas, 
is found in $U_\mathrm{ind}$ if there 
is no difference in the energy dispersion between the components, 
or simply $t_{\downarrow\alpha}=t_{\uparrow\alpha}$, which is the case that 
we mainly consider here. 

In the weak coupling regime, the induced interaction correction near the Fermi 
surface dominantly contributes to the calculations of the BCS order 
parameter and the critical temperature~\cite{Heiselberg2000a,Baranov2008a},  
and then the effective interaction is approximately given by only 
the Fermi surface momenta. Averaging $U_\mathrm{ind}(\mathbf{p},\mathbf{k})$ 
over the Fermi surface, the induced interaction becomes
$\langle U_\mathrm{ind} \rangle = \frac{1}{|S_\uparrow||S_\downarrow|}
\int_{S_\uparrow}dS_\mathbf{p}\int_{S_\downarrow}dS_\mathbf{k}
U_\mathrm{ind}(\mathbf{p},\mathbf{k}) $,
where $S_\sigma$ denotes the Fermi surface of the component $\sigma$ and 
$|S_\sigma|=\int_{S_\sigma}dS$ is the area. Finally,
the effective interaction is written as  
$U_\mathrm{eff} = U_0 + \langle U_\mathrm{ind} \rangle$.
This effective interaction replaces the interatomic interaction in calculations.  
Having other parameters fixed, $U_\mathrm{eff}$ becomes zero 
when $U_0$ equals to $U_c \equiv -U_0^2/\langle U_\mathrm{ind} \rangle$.
The susceptibility of an interacting gas is given as 
$\chi_0/(1+U_0\chi_0)$ within the random phase approximation, where
$\chi_0$ is the susceptibility of the noninteracting gas. 
At $U_0=U_c$, this quantity diverges, often indicating a possibility of
charge ordered phase. Our approach is formally valid 
below $U_c$, and with the GMB correction, the criterion for the weak coupling 
regime can be established by $U_c \ll U_0 < 0$.

We calculate zero-temperature order parameters by using this effective 
interaction $U_\mathrm{eff}$ 
in the mean-field formalism~\cite{Koponen2007a,Koponen2008a}.
The diagram in Fig.~\ref{fig1}(a) representing the induced interaction 
is relevant in normal states, and thus the critical temperature $T_c$ 
has been of interest in previous studies. However, in the weak 
coupling limit, the validity of the induced interaction can be readily 
extended for the calculation of the zero-temperature order parameter 
$\Delta$ because the contribution of the broken symmetry 
phase is expected to be of higher order in $\Delta$ that becomes 
very small.   
Moreover, with $U\equiv U_\mathrm{eff}$, we confirm that a general 
relation $2\Delta/k_BT_c \sim 3.53$~\cite{MartinRodero1992a,Pethick2008a} 
still holds in lattices, as shown in Fig.~\ref{fig1}(b), 
in the weak coupling regime. The calculation of $\Delta$ is 
computationally less demanding than $T_c$ but gives a concise picture 
of the transition.  
At very low filling factors (small $\mu$), one recovers the well-known 
prefactor $2.22$ in comparison between the usual BCS result and 
our calculation with the correction in $U_\mathrm{eff}$.  
In contrast, as the filling factor increases, the reduction 
of the order parameter with the correction becomes much more subtle
in a lattice than indicated by the prefactor in continuum. 

\begin{figure}
\includegraphics[width=0.47\textwidth]{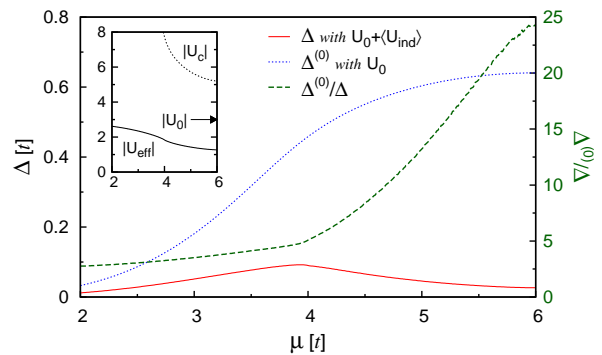} 
\caption{
\textit{Three-dimensional lattices.} 
The order parameters $\Delta$ (with the GMB correction) 
and $\Delta^{(0)}$ (the usual BCS result) are calculated with $U_0=-3t$ 
as a function of chemical potential $\mu$. 
The comparison between $\Delta$ and $\Delta^{(0)}$ shows an increasing 
deviation as $\mu$ increases. The ratio $\Delta^{(0)}/\Delta$ becomes 
$\sim 25$ at half filling ($\mu=6t$). The inset shows comparison 
between the magnitude of the interactions. Here and in the other figures, 
the critical coupling $|U_c|$ (see text) allows to estimate the ranges 
of $|U_0|$ and $\mu$ where the BCS mean-field theory with the GMB correction
is applicable. 
}
\label{fig2}
\end{figure}

Figure~\ref{fig2} shows the effect of the induced interaction
in isotropic 3D lattices. We find that 
the order parameter $\Delta$ with the correction shows a dramatic deviation 
from the usual BCS result $\Delta^{(0)}$ without the correction.
The ratio $\Delta^{(0)}/\Delta$ turns out to be nearly $25$ 
at half filling, which implies that the usual BCS prediction 
largely overestimates the order parameter and critical temperature. 
While the order parameter $\Delta$ with the correction is maximized around 
$\mu=4t$ where the Fermi surface reaches the Brillouin zone boundaries, 
it decreases substantially at higher filling factors
because of increasing contributions of states with 
$\mathbf{p}+\mathbf{k}+\mathbf{q}$ in Eq.~(\ref{eq:induced}) 
outside the first Brillouin zone.  
This deviation from the usual BCS result that we find here 
is qualitatively consistent with the previous study of the $1/D$ correction 
in high dimensions~\cite{vanDongen1991a},
where the order-of-unity reduction of the order parameter was estimated 
at half filling. However, our calculations in 3D lattices reveal 
much more significant suppression in $\Delta$ with the GMB correction.  
In all the figures, the curve 
for $|U_c|$ provides a simple estimate of the area in the parameter space 
($U_0$, $\mu$) where the GMB correction can extend the validity 
of the BCS mean-field theory.
The range of $U_0$ in this space decreases as $\mu$ increases towards
half filling in 3D and 2D lattices,which may affect the accuracy.  
Particularly in 2D, half filling is not included in this space 
because of divergent $\langle U_{\mathrm{ind}}\rangle$.

\begin{figure}
\includegraphics[width=0.47\textwidth]{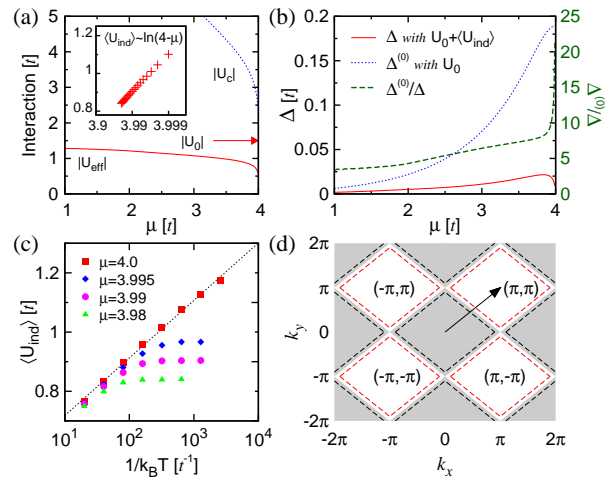} 
\caption{
\textit{Two-dimensional lattices.} 
(a) The effective interaction $U_\mathrm{eff}$  
and (b) the order parameter $\Delta$ are 
calculated with $U_0=-1.5t$. 
The usual BCS mean-field result $\Delta^{(0)}$ 
shows very large deviation from $\Delta$ with the GMB correction, 
exhibiting $\Delta^{(0)}/\Delta \approx 10$ near half filling $\mu=4t$. 
(c) Logarithmic divergence of the GMB correction and 
(d) Fermi seas (shaded area) at half filling indicating 
the nesting of Fermi surfaces (dashed lines) with momentum transfers 
$\mathbf{p}+\mathbf{k}=(\pm\pi,\pm\pi)$. 
}
\label{fig3}
\end{figure}

In 2D lattices, the order parameter $\Delta$ decreases very rapidly 
near half filling as the screening by the induced interaction dominates
(see Fig.~\ref{fig3}). This rapid decrease of $\Delta$ 
near half filling is in agreement with the previous QMC 
results on the critical temperatures~\cite{Scalettar1989a,Moreo1991a}.  
In contrast, the usual BCS mean-field calculation 
without the correction suggests a monotonically increasing 
order parameter $\Delta^{(0)}$ when approaching half filling.
At lower filling factors, the order parameter with the GMB correction 
turns out to be around $5$ times smaller than the one without the correction, 
which is consistent with other previous estimations of many-body 
effects~\cite{MartinRodero1992a,Deisz2002a,Strack2008a}. 
At higher $\mu$ close to half filling, the deviation from the usual BCS 
theory becomes even more substantial and leads to highly suppressed order
parameter near half filling. 

At half filling, it is known that the 2D attractive Hubbard model 
has the charge-density-wave order and the pairing order coexisting 
in the ground state, and the critical temperature of 
the superfluid transition goes to zero. 
Our calculations show that the induced interaction logarithmically 
diverges because of the Fermi surface nesting, making
the denominator in Eq.~(\ref{eq:induced}) infinitesimally small with 
the nesting vector $\mathbf{p}+\mathbf{k}=(\pm\pi,\pm\pi)$, 
for \textit{all} $\mathbf{q}$ at the Fermi surface, by mapping one side of 
the Fermi surface onto the other side [see Fig.~\ref{fig3}(d)].
Despite the tendency for the order parameter 
$\Delta$ to vanish when approaching half filling, 
the order parameter is not well defined with the large correction 
in our perturbative approach. However, this divergence of the induced 
interaction means that, for arbitrary small $U_0$, the
susceptibility diverges. It thus provides the connection to a different 
type of phase that cannot be anticipated by the usual BCS mean-field theory. 
The divergent Lindhard function appearing in the correction term can be 
interpreted as the signature of the charge density waves~\cite{Gruner1988a}.

For direct comparison with the QMC values of $T_c$ in 2D lattices, 
we have used $U_0=-4t$. With the GMB corrections, 
we obtained $\Delta~(T_c) \sim$ $0.07t~(0.04t)$, $0.03t~(0.02t)$, 
$0.008t~(0.005t)$ at the filling factors ($|U_c|$'s) of 
$0.2~(7.3t)$, $0.25~(6.2t)$, $0.3~(5.4t)$. 
The farther away $U_c$ is from $U_0$, the more accurate $\Delta$ is expected.
The QMC result $k_B T_c \sim 0.05t$ at quarter filling
~\cite{Scalettar1989a,Moreo1991a} is remarkably close to our value of $\Delta$, 
and the results in \cite{Paiva2004a} are of the same order 
of magnitude as ours. 
Note that the usual BCS mean-field calculations for these parameters 
would give results that are about 10-35 times larger than the
QMC and our GMB corrected mean-field results.   

\begin{figure}
\includegraphics[width=0.47\textwidth]{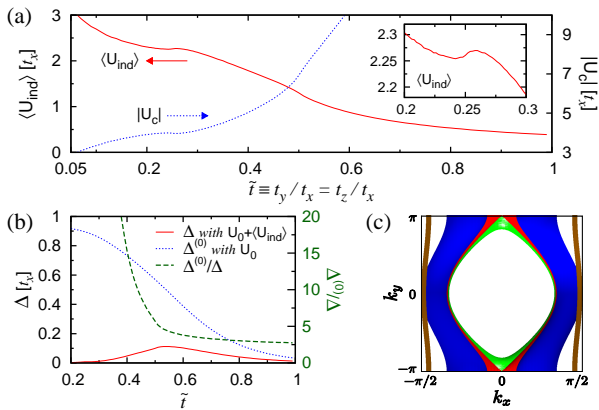}
\caption{
\textit{Crossover from 3D to 1D.}
(a) The induced interaction $\langle U_\mathrm{ind} \rangle$ and 
(b) the order parameter $\Delta$ as 
a function of the lattice anisotropy 
$\tilde{t} \equiv t_y/t_x = t_z/t_x$ in anisotropic three-dimensional lattices.
In (b), the usual BCS mean-field results $\Delta^{(0)}$ 
without the correction is given for comparison. 
(c) Fermi surfaces projected to $k_x$-$k_y$ 
space at $\tilde{t}=0.26,0.25,0.24,0.05$ (from center). 
The chemical potential is fixed at $\mu=2t_x$ and 
$U_0=-3t_x$ is used in the calculations. 
}
\label{fig4}
\end{figure}

Motivated by the fact that anisotropy is easily controllable 
in optical lattices, we now explore dimensional 
crossover from 3D to 1D by introducing directional difference 
in the tunneling strengths $t_\alpha$.
For this purpose, we define the lattice anisotropy as a ratio 
of the tunneling strengths, $\tilde{t} \equiv t_y/t_x = t_z/t_x$, 
with which one can change the dimensionality from 3D ($\tilde{t}=1$) 
to 1D ($\tilde{t}=0$). Figure~\ref{fig4} shows the effect of the lattice 
anisotropy on the induced interaction and the order parameter. 
As the anisotropy evolves with $\tilde{t}$, the screening by the induced 
interaction becomes stronger and finally diverges in the limit of 1D because of 
Fermi surface nesting. 

Similar to isotropic 3D cases, the order parameter $\Delta$ in the anisotropic
lattice also shows highly suppressed values compared with the usual BCS 
result $\Delta^{(0)}$. 
While the transition from 3D to 1D appears continuous in
the induced interaction $\langle U_\mathrm{ind} \rangle$, we identify   
two special points of $\tilde{t}$ indicating structural changes of 
the Fermi surface. 
First, $\langle U_\mathrm{ind} \rangle$ has a kink around $\tilde{t}=0.5$ 
at which $\Delta$ begins to decrease. The Fermi surface is closed originally 
in the 3D lattice with the given chemical potential $\mu=2t_x$. 
With decreasing $\tilde{t}$, the Fermi surface becomes deformed, 
and then at $\tilde{t}=0.5$, the Fermi surface becomes open.   
The second is a bump of $\langle U_\mathrm{ind} \rangle$ 
near $\tilde{t}=0.25$ at which a dimensional change of the Fermi surface 
occurs and the nesting effect develops to escalate 
$\langle U_\mathrm{ind} \rangle$.  As plotted in Fig.~\ref{fig4}(c), 
finally at $\tilde{t}=0.25$, the surface completely opens in $k_y$ and $k_z$ 
directions and splits into two disconnected sheets causing the nesting effect. 

However, in the quasi-1D regime, it turns out that the parameter space given
by $U_c$ does not cover the low $\tilde{t}$ region, 
and our calculation predicts a vanishing order parameter at low $\tilde{t}$, 
which deviates from previous rigorous studies of the Hubbard model. 
In the attractive Hubbard model in quasi-1D, the spin gap and the critical 
temperature are finite~\cite{Larkin1978a}. In the limit of 1D, 
the gap is still finite though the critical temperature goes to 
zero~\cite{Giamarchi2004a}. Singlet superfluidity dominates in 
the ground state, but there is no true long range order in the 1D Hubbard 
model~\cite{Essler2005a}.   

We have also considered the problem of the fermions in component-dependent 
lattice potentials~\cite{Cazalilla2005a} where
each component  experiences a different tunneling strength in a lattice.
This difference in tunneling in a lattice is analogous to unequal effective 
masses of Fermi gases in continuum. 
In 3D lattices, we have found that the screening effect 
of the induced interaction becomes stronger as the difference between 
the tunneling strengths increases, which agrees well with the results 
for homogeneous gases~\cite{Baranov2008a,Paananen2007a} where similarly 
the stronger screening effect at the larger mass imbalance was found. 

In conclusion, we have found that the presence of the optical lattices 
substantially strengthens the effect of the GMB correction 
on the BCS superfluidity. 
The consequent suppression of the order parameter is found to be 
much beyond the ratio $2.22$ predicted in homogeneous gases, 
which agrees with the estimations of other previous many-body correction 
studies at low filling factors. As the filling factor becomes higher, 
the inclusion of the correction becomes increasingly important. 
For instance, the order parameter turns out to be almost 25 times 
smaller with the correction than the usual BCS mean-field results 
at half filing in 3D lattices. 
Moreover, the behavior of the order parameter 
in 2D lattices shows excellent agreement with the previous QMC values.
Naturally, when the correction becomes very large, our 
perturbative approach breaks. The divergence of the correction 
is related to the phase at half filling in 2D where superfluid order and 
charge-density-wave order coexist.

One of the general shortcomings of a mean-field theory is that it gives 
a valid approximation only in high dimensions. Particularly for BCS 
superfluidity, our findings suggest that the effective theory with 
the many-body correction to the interatomic interaction can significantly 
extend the applicability of the mean-field calculations 
in the lower dimensions, namely 3D and 2D, and in the crossover from 3D to 
1D lattices, in spite of the obvious failure in the strict 1D limit.
With the GMB correction, the simple mean-field calculation can also provide 
quantitatively reliable values in a wider range of the coupling strength,
without sophisticated QMC calculations. 
The unanticipated large suppression of the order parameter 
at high filling factors highlights the practical importance of our results, 
which may provide a new insight to the issue of the critical temperature 
in future realizations of fermionic superfluids in optical lattices. 

\begin{acknowledgments}
The authors thank Dr.\ T.K.\ Koponen for fruitful discussions. 
This work was supported by Academy of Finland and EuroQUAM/FerMix
(Projects No. 213362, 217041, 217043, and 210953) and conducted
as a part of a EURYI scheme grant \cite{euryi}.
\end{acknowledgments}

\end{document}